\documentclass[aps,prl,reprint,floatfix]{revtex4-1}

\usepackage{natbib,graphicx,amssymb,amsmath,xcolor,bm,hyperref}

\begin{document}
\title[]{A way to generate poloidal (zonal)  flow in the dynamics of drift (Rossby) waves}

\author{Alexander M. Balk}
%\email[balk@math.utah.edu]
\affiliation{Department of Mathematics, University of Utah,
155 South 1400 East, Salt Lake City, Utah 84112}

\date{\today}

\begin{abstract}
The paper considers dynamics in the Charney-Hasegawa-Mima equation, basic to several different phenomena. 
In each of them, the generation of poloidal/zonal flow is important.  
The paper suggests a possibility to generate such flows (which can serve as transport barriers). 
Namely, one needs to create significant {\it increment} and {\it decrement} in neighborhoods 
of some wave vectors ${\bf k}_1$ and ${\bf k}_2$ (respectively) such that
(1)  $R_{{\bf k}_1}<R_{{\bf k}_2}$,
where $R_{\bf k}$ is the spectral density of the extra invariant ($I=\int R_{\bf k} E_{\bf k} d{\bf k}$ is the extra invariant, with $E_{\bf k}$ being the energy spectrum),
(2) $|{\bf k}_1|<|{\bf k}_2|$, and 
(3) ${\bf k}_1+{\bf k}_2$ is a poloidal/zonal wave vector.
These three conditions define quite narrow region. 
\end{abstract} 


\maketitle
%\renewcommand{\theequation}{\thesection.\arabic{equation}}
%\section{Introduction}
1. The quasi-geostrophic or Charney-Hasegawa-Mima (CHM) equation \cite{Charney48,HasegawaMima77}
\begin{eqnarray}\label{QGE}
 (1-\Delta)\psi_t + \psi_y = \psi_x \Delta\psi_y - \phi_y\Delta\psi_x
\end{eqnarray}
is a basic model for several different phenomena, in particular,
$1^\ast$ ocean dynamics \cite{Va},
$2^\ast$ tokamak plasmas \cite{DiamondItoh10,Horton},
and $3^\ast$ slow magneto-hydrodynamics in the ocean of the core \cite{Brag07} 
(in the latter case, instead of being the stream function, $\psi$ is the vertical component of the vector potential, so
$-\psi_y, \psi_x$ are the horizontal components of magnetic field \cite{Balk2014}).

In all these situations the generation of poloidal/zonal flow is important.
In the first case emerging zonal flow limits the meridional transport.
In the second case, poloidal flow limits the transport in the radial direction of a tokamak.
In the third case, instead of zonal flow, we have  generation of zonal magnetic field, important for dynamo theory.

Unlike cases $1^\ast$ and $3^\ast$, in the case $2^\ast$ we can actually change something;
and this is the reason why the CHM equation is written here in the plasma notations (with radial $x$ and poloidal $y$ coordinates;  units are chosen to make coefficients equal 1). 

The present paper describes what increment/decrement could we add to the CHM equation in order to generate or to aid in the generation of poloidal flow (a transport barrier).

In the Fourier representation, the equation (\ref{QGE}) --- with the increments $\gamma_{\bf k}$ --- becomes
 \begin{eqnarray}\label{Fourier}
\dot \psi_{\bf k}+i\Omega_{\bf k} \psi_{\bf k}=\frac{1}{2} \int W_{-{\bf k},\alpha,\beta} \psi_\alpha \psi_\beta \, d_{\alpha\,\beta} \,+\, \gamma_{\bf k} \psi_{\bf k},
\end{eqnarray}
where indices $\alpha,\beta$ (and later $\lambda,\mu,\nu$) stand for the corresponding wave vectors 
(${\bf k}_\alpha,\ldots$); $d_{\alpha\beta}=d{\bf k}_\alpha\, d{\bf k}_\beta$.
 \begin{eqnarray}\label{Dispersion}
 \Omega_{\bf k}=\frac{q}{1+k^2}
\end{eqnarray}
is the dispersion law [wave vector ${\bf k}=(p,q)$, $k^2=p^2+q^2$];
the coupling kernel
$W_{-{\bf k}\,\alpha\,\beta}=U_{-{\bf k}\,\alpha\,\beta}\;\delta(-{\bf k}+{\bf k}_\alpha+{\bf k}_\beta)$,
 \begin{eqnarray}\label{U}
 U_{-{\bf k}\,\alpha\,\beta}=(p_\alpha q_\beta - p_\beta q_\alpha)\frac{k_\beta^2-k_\alpha^2}{1+k^2}.
\end{eqnarray}

2. The equation (\ref{QGE}) is remarkable in the following sense \cite{BNZ,B1991}. 
In addition to the energy and the momentum (the enstrpophy is their linear combination), 
this equation has an (independent) extra invariant, conserved adiabatically, i.e.\ approximately over long time. 
To see this, consider the quantity
\begin{eqnarray}\label{I}
I=\frac{1}{2}\int X_{\bf k} |\psi_{\bf k}|^2\, d{\bf k}\;+\;
\frac{1}{6}\int Y_{\lambda\,\mu\,\nu}\psi_\lambda\,\psi_\mu\,\psi_\nu\, d_{\lambda\,\mu\,\nu}
\end{eqnarray}
with undetermined coefficient functions $X$ and $Y$ 
(without loss of generality, $Y$ is independent of the order of its indices, 
and $X$ is even, $X_{\bf k}=X_{-\bf k}$ since $\overline{\psi_{\bf k}}=\psi_{-{\bf k}}$). 
The time derivative of (\ref{I}) due to the equation (\ref{Fourier}) is 
\begin{eqnarray}\label{Idot}
\dot I&=&\int \gamma_{\bf k}\, X_{\bf k}\, |\psi_{\bf k}|^2\, d{\bf k}\nonumber\\
&+&\frac{1}{6} \int \{X_\lambda W_{\lambda\mu\nu} + X_\mu W_{\mu\nu\lambda} + X_\nu W_{\nu\lambda\mu}+
\nonumber\\ &&\hspace{-.5cm}
(\gamma_\lambda-i\Omega_\lambda+\gamma_\mu-i\Omega_\mu+\gamma_\nu-i\Omega_\nu)
 Y_{\lambda\mu\nu}\} \psi_\lambda\psi_\mu\psi_\nu\, d_{\lambda\mu\nu}\nonumber\\
 &+&\frac{1}{4}\int W_{-\lambda\alpha\beta} Y_{\lambda\mu\nu}
        \psi_\alpha \psi_\beta \psi_\mu \psi_\nu  d_{\alpha\beta\mu\nu}.
\end{eqnarray}
Suppose the drift waves have small amplitudes and small increments: 
$\psi_{\bf k}=O(\epsilon)$ and $\gamma_{\bf k}=O(\epsilon)$, $\epsilon$ is a small parameter.  
$O(\epsilon^3)$ terms in (\ref{Idot}) would cancel if we chose 
\begin{eqnarray}
Y_{\lambda\mu\nu}&=&
\frac{X_\lambda W_{\lambda\mu\nu} + X_\mu W_{\mu\nu\lambda} + X_\nu W_{\nu\lambda\mu}}
{i(\Omega_\lambda+\Omega_\mu+\Omega_\nu)};\label{Y}\\
\Longrightarrow\quad
\dot I&=&\int \gamma_{\bf k}\, X_{\bf k}\, |\psi_{\bf k}|^2\, d{\bf k}\;+\;O(\epsilon^4).\label{INVdot}
\end{eqnarray}

Without increment ($\gamma_{\bf k}\equiv 0$), we have $\dot I=O(\epsilon^4)$, while $I=O(\epsilon^2)$; integrating in time, we find $\Delta I\equiv I(t)-I(0)=O(\epsilon^3)$ over long time $t=O(\epsilon^{-1})$.
Notice that the $Y$-term in (\ref{I}) has the same order $O(\epsilon^3)$ as $\Delta I$, 
and so, the $Y$-term can be eventually discarded.

The above argument tacitly assumes that the expression (\ref{Y}) does not blow-up when its denominator vanishes. 
The  latter condition turns out to be very restrictive \cite{ZSch0}, realizable only for some special functions $X_{\bf k}$. 

If $X_{\bf k}=1+k^2$, then $Y\equiv 0$, and $I$ is the energy.

If $X_{\bf k}=k^2(1+k^2)$, then $Y\equiv 0$, and $I$ is the enstrophy.

There is one choice with non-zero $Y$: $X_{\bf k}=\frac{(1+k^2)^2}{q}\eta_{\bf k}$,
\begin{equation}\label{eta}
\eta_{\bf k}=\arctan\left(\frac{p+\sqrt{3}\, q}{k^2}\right)\,-\,\arctan\left(\frac{p-\sqrt{3}\, q}{k^2}\right)\,.
\end{equation}
The requirement of no blow-up in (\ref{Y}) is reduced \cite{BaVan} 
to the conservation  of function (\ref{eta}) in the 3-wave resonance interactions:
\begin{eqnarray}\label{ResonConserv}
\left. \begin{array}{c} {\bf k}_\lambda+{\bf k}_\mu+{\bf k}_\nu=0,\\ 
                                              \Omega_\lambda+\Omega_\mu+\Omega_\nu=0   \end{array}
\right\}\Rightarrow
\eta_\lambda+\eta_\mu+\eta_\nu=0
\end{eqnarray}
[recall that all three $W$ kernels in (\ref{Y}) contain the same delta function 
$\delta({\bf k}_\lambda+{\bf k}_\mu+{\bf k}_\nu)$]. The condition (\ref{ResonConserv}) {\it uniquely} \cite{BaFe} determines the function (\ref{eta}) --- up to linear combinations: 
Obviously, any linear combination of functions ${\bf k},\Omega_{\bf k},\eta_{\bf k}$ is also 
conserved in the 3-wave resonance interactions. Actually, it is beneficial, instead of function $\eta_{\bf k}$, to consider function $\tilde\eta_{\bf k}=\eta_{\bf k}-2\sqrt{3}\, \Omega_{\bf k}$. This combination vanishes as $k\rightarrow\infty$, faster than $\eta_{\bf k}$ and $\Omega_{\bf k}$ separately; and this takes place along all directions in the ${\bf k}$-plane. The function $\tilde\eta_{\bf k}$ gives a well defined invariant, that holds in the physical space as well \cite{BaYo}. The combination $\tilde\eta_{\bf k}$ also vanishes faster than $\eta_{\bf k}$ and $\Omega_{\bf k}$ separately, as $q\rightarrow 0$, for any $p$.

Thus, the equation (\ref{QGE}) has three invariants:
\begin{eqnarray}
&&\mbox{Energy }\quad E=\frac{1}{2}\int E_{\bf k}\, d{\bf k},\quad\dot E=\int \gamma_{\bf k} E_{\bf k}\, d{\bf k};
\label{En}\nonumber\\
&&\mbox{Enstrophy }\, \Phi=\frac{1}{2}\int k^2 E_{\bf k}\, d{\bf k},\;\dot\Phi=\int \gamma_{\bf k} k^2 E_{\bf k}\, d{\bf k};
\label{Es}\\
&&\mbox{Extra Invariant} \; I=\frac{1}{2}\int R_{\bf k} E_{\bf k}\, d{\bf k},\;\dot I=\int \gamma_{\bf k} R_{\bf k} E_{\bf k}\, d{\bf k},\label{Ex}\nonumber
\end{eqnarray}
\begin{eqnarray}\label{R}
R_{\bf k} =\frac{\eta_{\bf k}-2\sqrt{3}\,\Omega_{\bf k}}{\Omega_{\bf k}}\,.
\end{eqnarray}
We should keep in mind that the extra conservation holds only for weak nonlinearity, when $\epsilon$ is small enough, and the reminder $O(\epsilon^4)$ in (\ref{INVdot}) can be neglected.

3. All three invariants (\ref{Es}) are positive-definite. 

The presence of the extra invariant leads to essential conclusion about the energy transfer 
from the source (acting at some scale) to other scales \cite{BNZ}. 
The following argument \cite{B2005} describes the emergence of poloidal flow.

Due to the enstrophy conservation, the energy from the source would transfer towards larger scales, i.e.\ towards the origin in the $k$-plane. 
Due to the extra conservation, the energy should concentrate near the $p$-axis, that correspond to the poloidal  flow. 

\begin{figure}%[htb]
\includegraphics[width=\columnwidth]{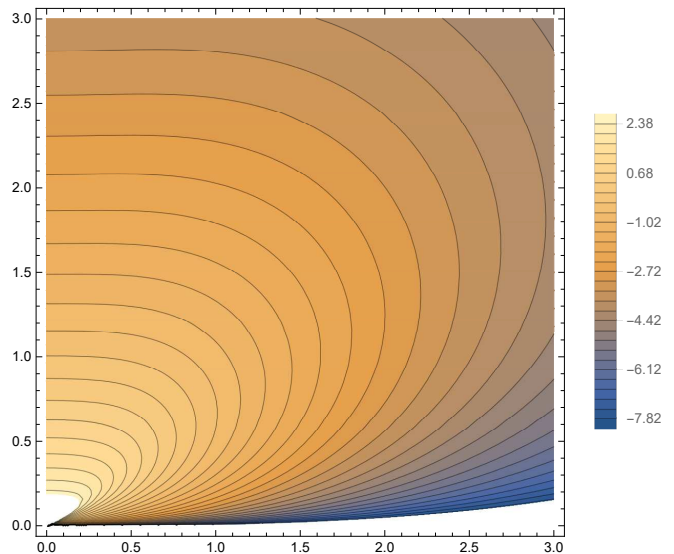}
\caption{\label{fig:Rcontour} Contour plot of the ratio (\ref{R}). 
The values of the color-bar are proportional to $\log_{10} R_{\bf k}$.
The white spots at the bottom correspond to too large values $|\log_{10} R_{\bf k}|$ (out of the range of the color-bar):
$R\sim\pi/q\rightarrow\infty$ as $k\rightarrow 0$, $|p/q|<\sqrt{3}$, 
and $R\propto q^2\rightarrow 0$ as $q\rightarrow 0$.}
\end{figure}

Indeed, Fig.\ \ref{fig:Rcontour} shows the contour plot of the ratio (\ref{R}). 
We see that $R_{\bf k}$ decreases when $k$ becomes large or when $q$ becomes small; more precisely,
\begin{eqnarray}\label{asy}
R_{\bf k} =8\sqrt{3}\times\left\{\begin{array}{lr}
\frac{5 p^2 q^2 + q^4}{5 k^8}\;+\;O(k^{-6}), & k\rightarrow\infty,\\
\frac{q^2}{p^2(1+p^2)^2}\;+\;O(q^4), & q\rightarrow\infty.
\end{array}\right.
\end{eqnarray}
So, if the energy from the source were transferred away from the $p$-axis, 
then the extra invariant would significantly increase:
From the right upper corner (big $k$) in Fig.\ \ref{fig:Rcontour} to the left bottom corner (small $k$) the ratio $R$ changes by 9 orders of magnitude, provided the small wave vectors belong to the sector of polar angles $\theta\equiv\arctan(q/p)>30^\circ$. 
When $\theta$ decreases from $30^\circ$ to $0^\circ$, the ratio $R$ decreases to zero. So, the difference in scales (between the source scale and the big scale) should be large enough, to ensure the energy concentration near the $p$-axis.

This reasoning equally applies to the cascade (local in the ${\bf k}$-plane) or non-local energy transfer.
So, the poloidal  flow is always generated (without us doing anything).

However, there is a problem with this argument: 
It is hardly possible in practice, that the drift waves are weakly nonlinear in a wide range of scales. Usually, smaller scales are strongly nonlinear, while larger scales are weakly nonlinear (e.g.\ \cite{Rhines}). Below we present an argument, showing the possibility of zonal flow generation without wide range of scales. But we do need to arrange increment/decrement in certain special way. 

This possibility is due to the fact that the contour lines in Fig.\ \ref{fig:Rcontour} have a ``depression'' in some region near the $q$-axis (away from the poloidal flow, corresponding to the $p$ axis): 
In this region, the ratio $R_{\bf k}$ is an {\it increasing} function of $p$ (while $q$ is fixed). 
This ``depression'' is hardly visible in Fig.\ \ref{fig:Rcontour}.

4. Let there be a positive increment in a small neighborhood $B_1$ of some wave vector ${\bf k}_1$ and a decrement in a small neighborhood $B_2$ of some ${\bf k}_2$, while $\gamma_{\bf k}\equiv0$ everywhere else. Consider positive quantities
\begin{eqnarray*}
G_1=\int_0^\infty dt \int_{B_1} \gamma_{\bf k}E_{\bf k} d{\bf k},\;
G_2=-\int_0^\infty dt \int_{B_2} \gamma_{\bf k}E_{\bf k} d{\bf k}.
\end{eqnarray*}
According to (\ref{Es}),
\begin{eqnarray*}
E^\star&=&E^0+G_1-G_2;\\
\Phi^\star&=&\Phi^0+k_1^2\,G_1 - k_2^2\,G_2;\\
I^\star&=&I^0+R_1\,G_1 - R_2\,G_2;
\end{eqnarray*}
here $E^0,\Phi^0,I^0$ are the initial values (at $t=0$) of the energy, enstrophy, extra invariant; 
their final values (at $t=\infty$) are $E^\star,\Phi^\star,I^\star$;\;
$R_j=R_{{\bf k}_j}\, (j=1,2)$. Slight difference between $k_1$ and $k_2$ (respectively, $R_1$ and $R_2$) could produce large difference between $\Phi^0$ and $\Phi^\star$ ($I^0$ and $I^\star$) if the time interval is long enough.

We want to generate significant amount of energy ($G_1>G_2$) and to have small amount of the extra invariant $I^\star\approx 0$. [Recall that the vanishing of the extra invariant requires vanishing of all drift waves besides the poloidal flow.] If $I^\star<I^0$, i.e.\ $R_1\,G_1-R_2\,G_2<0$, then 
\begin{eqnarray}\label{RR}
R_1<R_2\,.
\end{eqnarray}
The condition (\ref{RR}) is implied by even less stringent requirement that the generated extra invariant per generated energy  $\frac{R_1\,G_1 - R_2\,G_2}{G_1-G_2}$ is less than $R_1$ or $R_2$.

We want to generate {\it large-scale} flow. In other words, the energy should concentrate in longer waves. The latter carry less enstrophy per energy than shorter waves. Thus, we should pump mostly energy and dissipate mostly enstrophy, i.e.
\begin{eqnarray}\label{kk}
k_1<k_2\,.
\end{eqnarray}
Formally, the condition (\ref{kk}) follows from the requirement that the generated enstrophy per generated energy $\frac{k_1^2\,G_1 - k_2^2\,G_2}{G_1-G_2}$ is less than $k_1^2$ or $k_2^2$.

If (\ref{kk}) does not hold, $k_1>k_2$, then it is actually possible that the extra invariant becomes small, but the poloidal  flow is not generated; this can happen because the ratio $R_{\bf k}$ quickly decreases as $k$ increases [see Fig.\ \ref{fig:Rcontour} and asymptotics (\ref{asy})].

\begin{figure}%[htb]
\includegraphics[width=\columnwidth]{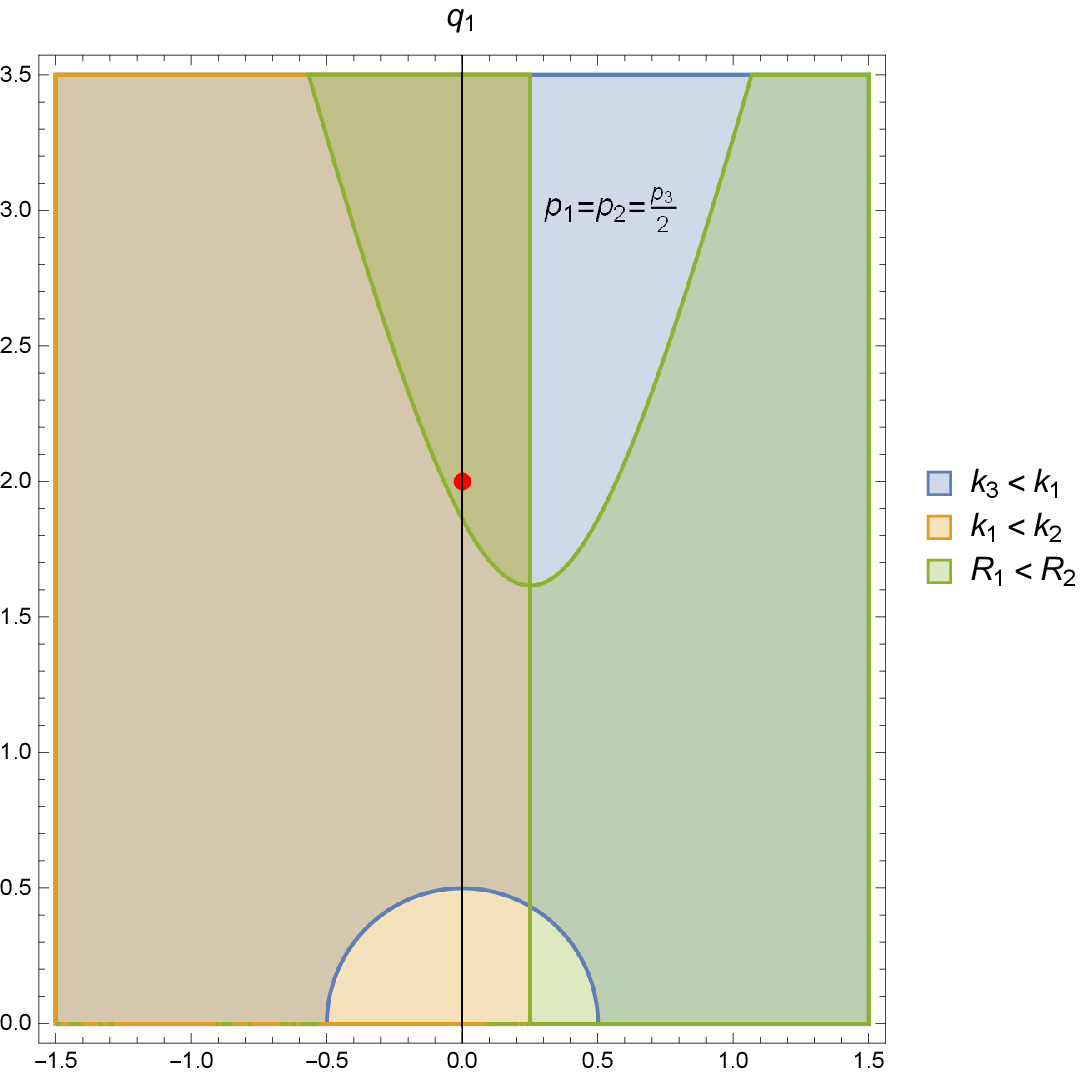}
\caption{\label{fig:p1q1} Three regions in the ${\bf k}_1$-plane (while $p_3=1/2$):\\
(i) outside of the semicircle is the region $k_3<k_1$;\\
(ii) left of vertical line $p_1=p_2=p_3/2$ is the region $k_1<k_2$;\\
(iii) the region $R_1<R_2$ consists of two pieces: 
right of the line and {\it outside} of the parabola-like curve, 
and left of the line and {\it inside} the parabola-like curve.\\
The intersection of regions (ii) and (iii) is the region satisfying the conditions (\ref{RR})-(\ref{pq}). 
Inside this intersection is the point ${\bf k}_1=(0,2)$ marked by the dot. 
(Due to the symmetry, only half of the domain, with $q_1>0$, is shown.)}
\end{figure}

We want to generate poloidal flow, and so, we require ${\bf k}_3={\bf k}_1+{\bf k}_2$ be a purely polodal wave vector: 
\begin{eqnarray}\label{pq}
p_2=p_3-p_1\quad q_2=-q_1\,.
\end{eqnarray}
Figure \ref{fig:p1q1} shows region of the ${\bf k}_1$-plane 
determined by the conditions (\ref{RR}) - (\ref{pq}), 
while $p_3$ is held fixed. These conditions automatically imply $k_3<k_1$. 
So,  the wave ${\bf k}_1$ --- pumped due to the positive increment --- can decay 
into the waves ${\bf k}_2$ and ${\bf k}_3$ \cite{Gill74}.

5.  There is an additional bonus of condition (\ref{kk}): It implies that one can make numerical simulation with only few modes, as significantly shorter waves are not generated. The generation of much longer waves is also impossible if the scales corresponding to $k_1,k_2$ are  close to the size of the system.

Using this, we consider the simplified dynamics of only three waves \cite{Balk18} with wave vectors 
${\bf k}_3=(1/2,0)$, 
${\bf k}_1=(0,2)$ corresponding to the dot in Fig.\ \ref{fig:p1q1}, 
and  ${\bf k}_2={\bf k}_3-{\bf k}_1$
\begin{eqnarray}\label{psi}
\dot\psi_1+i\Omega_1\psi_1&=&U_1 \psi_3\overline{\psi_2}+\gamma_1\psi_1,\nonumber\\ 
\dot\psi_2+i\Omega_2\psi_2&=&U_2 \psi_3\overline{\psi_1}+\gamma_2\psi_2,\\
\dot\psi_3+i\Omega_3\psi_3&=&U_3 \psi_1 \psi_2;\nonumber
\end{eqnarray}
$U_1=U_{{\bf k}_1, {\bf k}_3, -{\bf k}_2},\;
 U_2=U_{{\bf k}_2, {\bf k}_3, -{\bf k}_1},\;
 U_3=U_{{\bf k}_3, {\bf k}_1, {\bf k}_2},\\
\Omega_j=\Omega_{{\bf k}_j}\, (j=1,2,3),\;\;
\gamma_1>0,\;\gamma_2<0$. 

There is long history of modeling fusion plasmas by various small ode-systems (see \cite{MaRoCa,Rypina} and references cited therein).

\begin{figure}%[htb]
\includegraphics[width=\columnwidth]{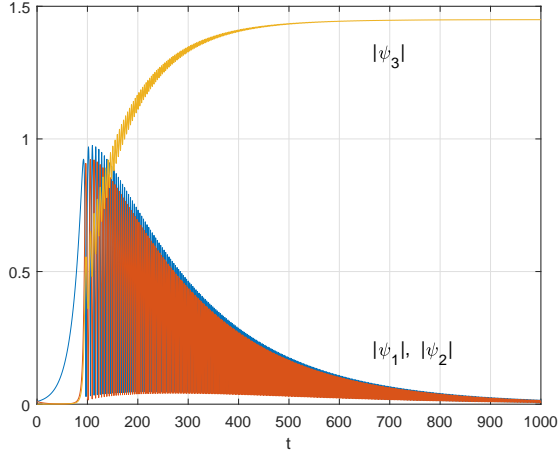}
\caption{\label{fig:simula}
Triad simulation.
The curves appear having some width; this is due to oscillations, cf.\ \cite{Balk18}. (Note, the figure shows oscillations of absolute values $|\psi|$, not $\psi$ themselves.)
For this particular graph, $\gamma_1=0.05,\,\gamma_2=-0.06;\;
|\psi_1(0)|=|\psi_2(0)|=0.01$ (the initial phases $\arg[\psi(0)]$ are random and appear insignificant).}
\end{figure}

Fig.\ \ref{fig:simula} shows the emergence of poloidal flow in the model (\ref{psi}).
I performed simulations using MATLAB ode-solvers 
with decreased absolute and relative tolerances (`AbsTol' \& `RelTol'). 
Instead of MATLAB default values `AbsTol'$=10^{-6}$ and 'RelTol'$=10^{-3}$, 
I used `AbsTol$=10^{-13}$ and 'RelTol'$=10^{-11}$. 
These are needed for validity of long-time simulations.
I also checked that different ode-solvers produced indistinguishable results.

The relevance of these calculations to tokamak plasmas is due 
to the smallness in seconds of the time scale for drift waves.
In particular, for ITER, the drift velocity is $v_d\sim2$km/s and the ion inertial (Rossby) radius $\rho\sim3$mm \cite{horton15iter}; 
therefore, the time scale is $\rho/v_d\sim(3/2)10^{-6}$s. So, dimensionless time $t=1000$ (the time-range in Fig.\  \ref{fig:simula}) corresponds to $1.5\times10^{-3}$s; within a fraction of this time, the amplitude of the poloidal mode becomes significantly bigger than the amplitudes of the other two modes. The emergence of the poloidal flow would occur faster if the increment and decrement  had bigger magnitudes.

6. If the emerging flow were not poloidal, it would be unstable with respect to decays into other waves. 
But poloidal  flow {\it is stable} in the weak interaction limit \cite{Gill74}. 
One can derive that the poloidal mode is stable if
\begin{eqnarray}\label{stability}
 p_3 |\psi_3|<1/(1+k_3^2),
\end{eqnarray}
i.e.\ the fluid velocity $v_3=i p_3 \psi_3$, due to the poloidal wave, 
has smaller magnitude than the phase velocity of this wave.
\footnote{Indeed, in the weak interaction limit, the instability of a Rossby wave ${\bf k}_3$ is the instability with respect to decays into a pair of waves, say ${\bf k}_4$ and ${\bf k}_5$, such that ${\bf k}_3={\bf k}_4+{\bf k}_5,$ and $k_3$ is in between $k_4$ and $k_5$ \cite{Gill74}. It is well known, that this instability does not take place if 
\begin{eqnarray*}
D\equiv\omega^2/4 - U_4 U_5 |\psi_3|^2 >0,
\end{eqnarray*}
where $\omega=\Omega_4+\Omega_5\; (\Omega_3=0)$, and according to (\ref{U}),
\begin{eqnarray*}
U_4=U({\bf k}_4,{\bf k}_3,-{\bf k}_5)=p_3 q_5 (k_3^2-k_5^2)/(1+k_4^2),\\
U_5=U({\bf k}_5,{\bf k}_3,-{\bf k}_4)=p_3 q_4 (k_3^2-k_4^2)/(1+k_5^2).
\end{eqnarray*}
Since $q_4+q_5=0$, 
\begin{eqnarray*}
\omega=q_4 \left( \frac{1}{1+k_4^2} - \frac{1}{1+k_3^2} \right)
           +q_5 \left( \frac{1}{1+k_5^2} - \frac{1}{1+k_3^2} \right).
\end{eqnarray*}
By inequality between arithmetic and geometric means,
\begin{eqnarray*}
\left(\frac{\omega}{2}\right)^2&\ge& q_4 q_5 
\frac{(k_3^2-k_4^2)(k_3^2-k_5^2)}{(1+k_4^2)(1+k_5^2)(1+k_3^2)^2},
\quad\Longrightarrow\\
D&\ge& q_4^2 \frac{(k_4^2-k_3^2)(k_3^2-k_5^2)}{(1+k_4^2)(1+k_5^2)}
\left\{\frac{1}{(1+k_3^2)^2}-p_3^2|\psi_3|^2\right\}.
\end{eqnarray*}
So, $D>0$ if the condition (\ref{stability}) holds.}

This stability in the weak interaction limit matches the argument based on the extra invariant that requires weak nonlinearity, as well. 

The considerations in the present paper can be extended to the dynamics in different equations; it is only important that such an equation has dispersion law (\ref{Dispersion}) and possesses Hamiltonian structure, e.g.\ see \cite{MajdaQi19}.

\bibliography{My}

\end{document}